\documentclass[nojss]{jss}

\usepackage{thumbpdf,lmodern} 

\usepackage{framed}
\usepackage{comment}
\usepackage{amsmath, calc}
\newcommand{\class}[1]{`\code{#1}'}
\newcommand{\fct}[1]{\code{#1()}}

\usepackage{Sweave} 

\author{Fabio Mason\\University of Geneva \And 
        Manuel Koller\\ \And
        Eva Cantoni\\University of Geneva \And 
        Paolo Ghisletta\\University of Geneva}

\Plainauthor{Fabio Mason, Manuel Koller, Eva Cantoni, Paolo Ghisletta}

\title{\pkg{confintROB}: An \proglang{R} Package for Confidence Intervals in Robust Linear Mixed Models}
\Plaintitle{confintROB Package: Confindence Intervals in robust linear mixed models}
\Shorttitle{confintROB Package}

\Abstract{
Statistical inference is a major scientific endeavor for many researchers. In terms of inferential methods implemented to mixed-effects models, significant progress has been made in the \proglang{R} software. However, these advances primarily concern classical estimators (ML, REML) and mainly focus on fixed effects. In the \pkg{confintROB} package, we have implemented various bootstrap methods for computing confidence intervals (CIs) not only for fixed effects but also for variance components. These methods can be implemented with the widely used \fct{lmer} function from the \pkg{lme4} package, as well as with the \fct{rlmer} function from the \pkg{robustlmm} package and the \fct{varComprob} function from the \pkg{robustvarComp} package. These functions implement robust estimation methods suitable for data with outliers. The \pkg{confintROB} package implements the Wald method for fixed effects, whereas for both fixed effects and variance components, two bootstrap methods are implemented: the parametric bootstrap and the wild bootstrap. Moreover, the \pkg{confintROB} package can obtain both the percentile and the bias-corrected accelerated versions of CIs.
}

\Plainkeywords{robust estimators; bootstrap; confidence intervals; linear mixed models; lme4; robustlmm; robustvarComp, R} 

\Address{
  Fabio Mason\\
  Faculty of Psychology and Educational Sciences\\
  University of Geneva\\
  Boulevard du Pont d'Arve 40, 1211 Geneva, Switzerland\\
  E-mail: \email{fabio.mason@unige.ch}\\\\
  Paolo Ghisletta\\
  Faculty of Psychology and Educational Sciences\\
  University of Geneva\\
  Boulevard du Pont d'Arve 40, 1211 Geneva, Switzerland\\
  E-mail: \email{paolo.ghisletta@unige.ch} \\
  \\\\
  Eva Cantoni\\
  Research Center for Statistics and Geneva School of Economics and Management\\
  University of Geneva\\
  Boulevard du Pont d'Arve 40, 1211 Geneva, Switzerland\\
  E-mail: \email{eva.cantoni@unige.ch}  \\\\
  Manuel Koller\\
  Zürich, Switzerland\\
  E-mail: \email{kollerma@protonmail.com}\\
}

\begin{document}
\Sconcordance{concordance:article_arXiv.tex:article_arXiv.Rnw:1 23 1 1 17 359 1 1 2 1 %
0 2 1 11 0 1 2 4 1 1 45 1 1 2 2 5 1 1 2 1 0 1 2 4 0 1 2 1 1 1 2 1 0 1 2 %
4 0 1 2 2 1 1 2 4 0 1 5 31 0 1 2 2 1 1 2 1 0 1 2 4 0 1 2 2 1 1 2 18 0 1 %
2 4 1 1 2 1 0 1 3 2 0 1 1 17 0 1 2 6 1 1 2 12 0 1 2 2 1 1 38 1 1 2 2 5 %
1 1 2 4 0 1 2 1 3 5 0 1 2 4 0 1 4 31 0 1 2 2 1 1 2 1 0 1 3 2 0 1 1 17 0 %
2 2 1 0 1 3 2 0 1 1 17 0 1 2 59 1}



\section[Introduction]{Introduction} \label{sec:intro}

The linear mixed model (LMM) is frequently applied to correlated multilevel data. The package \pkg{lme4} \citep{bates2015lme4} has notably contributed to the dissemination of LMM in various research fields. Indeed, it allows the analysis of data of variable degrees of complexity (e.g., hierarchical, cross-effects or mixed), with or without missing data, allows to compute confidence intervals (CIs) for the different types of parameter via the \fct{confint.merMod} function, and performs testing (and compute corresponding $p$-values) with the \pkg{lmerTest} package \citep{kuznetsova2017}. 

Recently, the field of robustness in the context of LMM has undergone significant developments \citep[e.g.][]{Pinheiro2001,copt2006,Chervoneva2014,Koller2013,agostinelli2016,saraceno2023,lucadamo2021,zhang2020,burger2018,mason2021,mason2024}. However, only a few of these proposals are implemented in \proglang{R}: the Robust Scoring Equations (RSE) estimator of \cite{Koller2013} and \cite{koller2023} is implemented in the package \pkg{robustlmm} \citep{Koller2016}, the composite-$\tau$ estimator (cTAU) of \cite{agostinelli2016} and the S-estimator (S) of \cite{copt2006} are implemented in the package \pkg{robustvarComp} \citep{agostinelli2016} and the multivariate $t$ maximum likelihood estimator (tML) of \cite{Pinheiro2001} is implemented in the \pkg{heavy} package \citep{Osorio2019}. For all these estimators, the Wald test (\fct{heavyLme} and \fct{varComprob} return $p$-values in the summary, whereas \fct{rlmer} returns standard errors) is the only statistical inference option, and with the exception of the estimators implemented in \pkg{robustvarComp}, this test only applies to the fixed effects. Unfortunately, the robust estimators are not compatible with the \pkg{lmerTest} package or the \fct{confint.merMod} function. This gap in existing software inspired us to create the \pkg{confintROB} package, which allows computing confidence intervals (CIs) based on the parametric and the wild bootstrap for both fixed effects and variance components of models estimated with the \fct{lmer} function from \pkg{lme4}, the \fct{rlmer} function from \pkg{robustlmm}, and with the \fct{varComprob} function from \pkg{robustvarComp}. Our package was inspired by the \fct{confint.merMod} and \fct{confint} functions for the arguments and for the output, with the difference that it implements two bootstrap methods also applicable to the robust estimators implemented in the \pkg{robustlmm} package and \pkg{robustvarComp} package, as well as the Wald type confidence intervals.

This article is intended as a tutorial for \fct{confintROB}, a function to compute CI with robust estimators implemented in the \proglang{R} environment \citep{singmann2015} in the homonymous package. The function can currently be applied to 5 classes of estimators and allows one to choose between 5 choices of CI types, including 4 bootstrap methods. The package is available on CRAN at \url{https://cran.r-project.org/package=confintROB} under the \href{https://cran.r-project.org/web/licenses/GPL-2}{GPL-2} license. 

In the next Section (2), we introduce the technicalities germane to the model and the estimators, and provide details on the wild and the parametric bootstrap schemes in their \emph{percentile} and \emph{Bias-corrected and accelerated (BCa)} versions. In Section 3, we introduce a real data application with the function \fct{rlmer}. Limitation and future extensions are discussed in Section 4.


\section{Technicalities}
\subsection{The Linear Mixed Model (LMM) and its (robust) estimation}

The LMM can be written as:
\begin{equation}
\label{LMM_eq}
\begin{array}{cc}
\boldsymbol{y}_{i} = X_i \boldsymbol{\gamma} + Z_i \boldsymbol{b}_i + \boldsymbol{\varepsilon}_i, 
\end{array}
\end{equation}
where $\boldsymbol{y}_i$ is a vector of length $J_i$ containing the responses of cluster $i$ (e.g. participants in a longitudinal study), $\boldsymbol{\gamma}$ is a vector of coefficients for the $\emph{p}$ fixed effects, $X_i$ is the ($J_i$ $\times$ $p$) design matrix for fixed effects, $\boldsymbol{b}_i$ is the vector of random effects of length $\emph{q}$, independent of the errors $\boldsymbol{\varepsilon}_i$  and $Z_i$ is the ($J_i$ $\times$ $q$) design matrix for the random effects. The vector $\boldsymbol{\varepsilon}_i =  (\varepsilon_{i1}, \ldots, \varepsilon_{iJ_i})^T$ contains the $J_i$ individual error terms for cluster $i$. Generally, it is assumed that random effects and error terms are normally distributed around zero \citep{laird1982}:
\begin{equation}\label{Nassump}
\boldsymbol{b}_i \sim \mathcal{N}(\boldsymbol{0},\Sigma) \ \ \mbox{ and } \ \  \boldsymbol{\varepsilon}_i \sim \mathcal{N}(\boldsymbol{0},\sigma_{\varepsilon}^2 I),
\end{equation}
where $\Sigma = \Sigma(\boldsymbol{\theta})$ is the $q \times q$ (parameterized) covariance matrix of the random effects and $\sigma_{\varepsilon}^2 I$ is the $(J_i \times J_i)$ (diagonal) covariance matrix of the error terms. That is, the model assumes that the random effects $\boldsymbol{b}_i$ and errors $\boldsymbol{\varepsilon}_i$ stem from zero-centered normal distributions.
The entire set of parameters to be estimated  is $(\gamma, \sigma_{\varepsilon}^2, \theta)^T$. We consider two general classes of estimators: \emph{classical} and \emph{robust}. The classical estimators are based on the likelihood of observing the data, assuming the model to be correct in the population. The full maximum likelihood (ML) approach and the restricted maximum likelihood (REML) approach assume that both random effects and error terms are normally distributed (see Equation~\eqref{Nassump}). Because classical estimators may lead to biased estimates and wrong inferential conclusions with contaminated data (i.e., data that contain outliers, defined as observations with extremely low probability of occurrence assuming normality of random effects and errors terms as in Equation~\eqref{Nassump}; cf. \citet{mason2021,mason2024}), robust estimators have been developed \citep[e.g.][]{copt2006,agostinelli2016,Koller2016}. These estimators are thus constructed to reduce bias in estimation of parameters in the presence of outliers.

\subsubsection{Classical Estimation}

The classical estimator of the LMM is maximum likelihood (ML), which assumes multivariate normality \citep[e.g.][]{richardson1995,welsh1997}. The (log-)likelihood function expresses the likelihood of the data as a function of the model parameters, assuming the model is correct. The estimates are the parameter values that maximize this function, for both fixed effects $\boldsymbol{\gamma}$ and variance components $\sigma_{\varepsilon}^2$ and $\boldsymbol{\theta}$. \\

The ML can be implemented in two versions, either full (ML), or restricted (REML). The main difference is that the latter estimates the variance components contingent on the fixed effects, while the former estimates both sets of parameters simultaneously. The function \fct{lmer} from the package \pkg{lme4} \citep{bates2015lme4} in \proglang{R} implements both versions.\\

\subsubsection{Robust Estimation}

Several robust methods have been proposed for the LMM. The \pkg{confintROB} package is available with two functions that implement a total of 8 robust estimators. The \fct{rlmer} function implements the RSE estimator and its fast approximative version (fast RSE), whereas the \fct{varComprob} function implements the $\tau$-, S- and MM-estimators and their \emph{composite} versions (see Table~\ref{table_est}). All estimators presume a so-called \textit{central model} \citep*[see, e.g.,][]{Koller2013}, where a central model $G$ is assumed for the majority of data (e.g., a normal distribution) and an alternative unknown model $H$ is assumed for the outliers, which are hence assumed from a different population. Typically, the substantive focus of the analysis is on the parameters of $G$. All 8 robust estimators also provide robustness weights that are applied to the data during the estimation procedure. The weights reveal important information about the possible outliers, because they indicate the extent to which the data are supposed to stem from the $G$ (weights close to $1$) rather than the $H$ (weights close to $0$) model. Whereas some robust methods estimate weights at the single observation level, to accommodate within-subject outliers (i.e., outlying points within the vector of an otherwise non-outlying subject), others only estimate subject-specific weights, which are applied to all observations of a given subject and are thus ideal for identifying between-subject outliers (i.e., subjects whose entire vector of data are atypical for $G$). Other methods yet estimate simultaneously both types of weights to accommodate both within- and between-subject outliers.\\

In this paper we discuss the RSE estimator implemented in the \fct{rlmer} function. This function allows to handle data with complex structure: nested or non-nested observations, balanced or unbalanced designs, with or without missing data, correlated or non-correlated random effects, etc.
\begin{table}[t!]
\centering
\begin{tabular}{llllp{5.4cm}}
\hline
Type           & Function & Package & Estimator & Argument\\ \hline
Classical         & \fct{lmer} & \pkg{lme4}       & ML  &    \code{REML = FALSE} \\
              &       &       & REML & \code{REML = TRUE}\\\hline
Robust & \fct{rlmer}   & \pkg{robustlmm}    &  RSE & \code{method = "DAStau"}\\
 &       &     & fast RSE & \code{method = "DASvar"}\\
               & \fct{varComprob}   & \pkg{robustvarComp}        & $\tau$ & \code{method = "Tau"}\\ 
                &       &     & composite-$\tau$ & \code{method = "compositeTau"}\\               
  &        &    & S & \code{method = "S"}\\
   &       &     & composite-S & \code{method = "compositeS"}\\
   &        &    & MM & \code{method = "MM"}\\
   &        &    & composite-MM & \code{method = "compositeMM"}\\\hline
\end{tabular}
\caption{\label{table_est} Overview of available functions and estimators with \fct{confintROB}.}
\end{table}

\subsection{Statistical Inference in the LMM}

In the frequentist approach to statistical inference, assessing a null hypothesis is usually performed by either performing an appropriate statistical test about a parameter of interest and comparing its $p$-value to a given statistical significance level, or by building a CI for that parameter and checking whether it covers the value 0. 

\subsubsection{Tests}\label{pval}
For LMM, \cite{kuznetsova2017} developed the \pkg{lmerTest} package that implements F-, and $t$-tests for the fixed effects parameters with degrees of freedom calculated following the Satterthwaite \citep{satterthwaite1946} or the Kenward-Roger \citep{Kenward_Roger_1997} approximations. For the fixed effect parameters, these tests are largely considered as the benchmark for classical LMM estimators \citep{kuznetsova2017}. The package also implements the likelihood ratio test (LRT) for the random effects. These methods are applicable only to the classical LMM estimators and objects of class \class{lmerMod} obtained via the \fct{lmer} function.

With robust estimators, standard errors are computed for the fixed effects with \fct{rlmer} and \fct{varComprob}, and for random effects with the latter only. With these quantities, one can then construct $t$-test statistics and compute associated $p$-values according to the reference distribution (Student $t$).
\\

\subsubsection{Confidence Intervals}
Confidence Intervals (CIs) are an alternative approach to perform statistical inference. The \pkg{lme4} package implements three kinds of CI for objects of class \class{lmerMod} with the \\ \fct{confint.merMod} function: the \emph{profile} method based on the LRT test, which is available only with ML estimates, the \emph{Wald} method based on the standard error estimates, and available only for fixed effects, and the \emph{bootstrap} method, which is often recommended to obtain CIs for fixed effects and variance components for LMM parameters \citep[e.g., ][]{Koller2016,Modugno2013}. 

To obtain bootstrap CIs, first, the empirical distribution of the parameter of interest, let's say $\beta$, is constructed, based on the multiple estimations $\hat{\beta}_1^*, \ldots, \hat{\beta}_B^*$ of the parameter in the $B$ resamples. With the percentile version, implemented in \fct{confint.merMod}, the lower and upper bounds are obtained from the $\frac{\alpha}{2}\times100$ and $(1-\frac{\alpha}{2})\times100$ empirical percentiles of the bootstrap distribution. When the bootstrap distribution is skewed and/or biased, \cite{tibshirani1993} proposed a refined version, called bias-corrected and accelerated (BCa), which is however computationally very intensive. To mitigate potential bias, the BCa estimates the parameter $\widehat{z}_0$ that corresponds to the proportion of estimates from the resamples smaller than the estimate of interest from the original sample. To correct for skewness, the ``accelerated'' parameter $\widehat{a}$ is estimated using the \textit{Jacknife} procedure \citep[e.g.][]{efron_bootstrap_1979}, resulting in

\begin{equation*}\label{BCa_a}
        \widehat{a} = \frac{\sum_{i=1}^n \left(\widehat{\beta}_{(\cdot)}-\widehat{\beta}_{(i)}\right)^3}{6 \left[\sum_{i=1}^n \left(\widehat{\beta}_{(\cdot)}-\widehat{\beta}_{(i)}\right)^2 \right]^{3/2}},
    \end{equation*}
    
    where $n$ is the number of subjects, $\widehat{\beta}_{(i)}$ is the estimate of $\beta$ on the sample without subject $i$ and $\widehat{\beta}_{(\cdot)}$ is the mean of the $n$ $\widehat{\beta}_{(i)}$ obtained. 
The BCa interval is then constructed as follows:

\begin{equation*}\label{BCa_IC}
        [\widehat{\beta}^*_{\alpha1} ; \widehat{\beta}^*_{\alpha2}], 
    \end{equation*}
    
    where
    
    \begin{equation*}\label{BCa_alp}
{\alpha1} = \Phi\left(\widehat{z}_0 + \frac{\widehat{z}_0 + z_{\alpha/2}}{1-\widehat{a}(\widehat{z}_0 + z_{\alpha/2})}\right)
 \end{equation*} 
 
 and
 
 \begin{equation*}\label{BCa_alpha2}
{\alpha2} = \Phi\left(\widehat{z}_0 + \frac{\widehat{z}_0 + z_{1-\alpha/2}}{1-\widehat{a}(\widehat{z}_0 + z_{1-\alpha/2})}\right),
 \end{equation*}
 
where $z_{\alpha/2}$ and $z_{1-\alpha/2}$ are the $\alpha/2$ and $(1-\alpha/2)$ quantiles of a Gaussian distribution respectively, $\alpha$ is the level of confidence, and $\Phi(\cdot)$ is the cumulative distribution function of a Gaussian distribution.

The \fct{confintROB} function implements two kinds of bootstrap (parametric and wild) in both versions (percentile and BCa), plus the Wald-type confidence interval, for a total of 5 options available with objects from \fct{lmer}, \fct{rlmer} and \fct{varComprob}. All methods are summarized in Table~\ref{table_ci}. \\

\paragraph{Parametric Bootstrap} First, the procedure estimates fixed effects ($\boldsymbol{\gamma}$) and variance component parameters $(\sigma_{\varepsilon}^2, \boldsymbol{\theta})$ from model~\eqref{LMM_eq} and~\eqref{Nassump} on the original data $(\boldsymbol{y}_i, X_i, Z_i ; i=1, \ldots,n)$, to produce $\hat{\boldsymbol{\gamma}}$, $\hat{\sigma}_{\varepsilon}^2$, and $\Sigma(\hat{\boldsymbol{\theta}})$. Then, for $b=1, \ldots,B$, the procedure :\\
\label{TOL_param.algo.page}
\begin{enumerate}

    \item \label{Tol_pb-1} for $i=1,\ldots, n,$ generates $\boldsymbol{\varepsilon}_i^{*} \sim \mathcal{N}(\hat{\sigma}_{\varepsilon}^2)$ and $\boldsymbol{b}_i^{*} \sim \mathcal{N}(\Sigma(\hat{\boldsymbol{\theta}}))$ to build
    \begin{equation*}\label{TOLparamB}
        \boldsymbol{y}_i^{*}= X_i \widehat{\boldsymbol{\gamma}} + Z_i \boldsymbol{b}_i^{*} + \boldsymbol{\varepsilon}_i^{*}
    \end{equation*}

    \item \label{Tol_pb-2} estimates $\boldsymbol{\gamma}$, $\sigma_{\varepsilon}^2$, and $\boldsymbol{\theta}$ from model \eqref{LMM_eq} and~\eqref{Nassump} on the resample $(\boldsymbol{y}_i^{*}, X_i, Z_i ; i=1, \ldots,n)$ to produce $\hat{\boldsymbol{\gamma}}^*$, $(\hat{\sigma}_{\varepsilon}^2)^*$, and $\Sigma(\hat{\boldsymbol{\theta}}^*)$.
\end{enumerate}

The estimator used at step~\ref{Tol_pb-2} can either be the same as, or differ from, that used on the original sample. Because usually $B=5000$ and because robust LMM estimation can be computationally very heavy\footnote{\label{Tol_comptime-fn} Over 500 large simulated samples (40 subjects measured 80 times following the design described in \cite{Modugno2013}) the mean computing time for a CI is 3.034 days with the first method and 0.9 days with the second method on the most powerful computing facility at the University of Geneva (4300 processors, 50 GPUs, up to 1.5TB RAM per calculation server, 1.2Po of dedicated storage).}, we choose to apply ML at step 2 also when a robust estimator was used at step 1 (cf. \cite{mason2024}).

\paragraph{Wild Bootstrap}The wild bootstrap method replaces step~\ref{Tol_pb-1} of the parametric bootstrap  computing ``residuals'' $\tilde{\boldsymbol{\upsilon}}_i$ as follows \citep{Modugno2013}:

\begin{equation*}\label{TOL_wildREStilde}
\tilde{\boldsymbol{\upsilon}}_i=\mbox{diag}(I - H_i)^{-1/2} \circ (\boldsymbol{y}_i - X_i\widehat{\boldsymbol{\gamma}}),
\end{equation*}

with $H_i=X_i(X^{T}X)^{-1}X_i^{T}$, where the operator ``$\circ$'' denotes the (element-wise) Hadamard product. \\

For $i=1, \ldots, n$, the procedure samples independently $w^{*}_i$  from the following standardized distribution \citep{Mammen1993}:

\begin{equation*}\label{TOL_w}
w^{*}_i=\left\{
\begin{array}{rl}
     - \dfrac{\sqrt{5}-1}{2} &  \mbox{with probability} \ p = (\sqrt{5}+1)/(2\sqrt{5})\\
    \dfrac{\sqrt{5}+1}{2}  & \mbox{with probability} \ q = 1-p
\end{array}  \right.
\end{equation*}

and builds individual responses 

\begin{equation*}\label{TOL_wildB}
\boldsymbol{y}^{*}_i= X_i \widehat{\boldsymbol{\gamma}} + \tilde{\boldsymbol{\upsilon}_i} w^{*}_i.
\end{equation*}

The wild bootstrap can be applied to all LMM estimators considered here.\\

\citet{Modugno2013} compared several methods to generate the resamples, including the parametric bootstrap and the semi-parametric wild bootstrap. The assumptions are stronger for the former method than for the latter (i.e., fixed covariates, correct specification of both fixed effects and variance components, and homoscedasticity and normality for random effects). The authors found that the semi-parametric wild method was superior in coverage in large samples and with heteroscedastic random effects. \\

\begin{table}[t!]
\centering
\begin{tabular}{lllp{7.4cm}}
\hline
Method    & Type  & Estimates     & Function specification \\ \hline
Wald    & - & Fixed effects   & \code{confintROB(object = model.output, method = "Wald")}       \\
Percentile   & wild & All  & \code{confintROB(object = model.output, boot.type = "wild")}  \\
  & parametric & All  & \code{confintROB(object = model.output, boot.type = "parametric")}  \\
BCa   & wild & All  & \code{confintROB(object = model.output, clusterID = "id", method = "BCa", boot.type = "wild")}  \\
   & parametric & All  & \code{confintROB(object = model.output, clusterID = "id", method = "BCa", boot.type = "parametric")}  \\ \hline
\end{tabular}
\caption{\label{table_ci} Overview of available methods to compute CI with \fct{confintROB}. }
\end{table}



\subsection{Design of the confintROB function}

The structure of the \fct{confintROB} function is inspired by the \fct{confint.merMod} function from the \pkg{lme4} package, which computes CIs from objects of class \class{lmerMod}. Similarly to \fct{confint.merMod}, the value of a \code{confintROB} object is a matrix containing the names of parameters in the first column, the lower and upper bounds of CIs in the second and third column, respectively, labelled with the corresponding percentages. The argument \code{parm} can be used to select a subset of parameters of interest using a vector of string characters indicating their names, or numerical values indicating the corresponding rows in the full matrix. The first argument is \code{object}, the fitted model with either \fct{lmer} (as \fct{confint.merMod}), \fct{rlmer} or \fct{varComprob}. The type of CI is chosen with the argument \code{method}, which can take the values \code{"Wald"}, \code{"boot"} (default), or \code{"BCa"}, which is an addition to \fct{confint.merMod}. If \code{"boot"} or \code{"BCa"} are selected, then argument \code{boot.type} needs to be specified, choosing from \code{"wild"} (default) or \code{"parametric"}. The argument \code{clusterID} is needed only with \code{method="BCa"} to specify the cluster variable with a string character corresponding to the variable name in the dataset. In the example of Section~\ref{illustration.section}, participants are the clusters, and the variable name that identifies the cluster is \code{"id"}. The numerical argument \code{level} is required to define the level of confidence and must be positive and smaller to 1, with default value .95. For bootstrap CIs (i.e., with \code{method = "boot"} or \code{method = "BCa"}), the number of bootstrap samples can be changed (the default value is 5000) with the argument \code{nsim}. There is an optional argument \code{verify.saved} that allows checking for a correspondence between the current results and those obtained, for example, in a previous version of a package in dependency. It requires an object produced by the \fct{confintROB} function.

\section{Illustration}
\label{illustration.section}

Here, we analyze the \code{medication} dataset, often used for didactic purposes \citep{Singer2003}, and originally discussed in \citet{tomarken1997}. The dataset is included in the package and contains 1242 observations with 5 variables:
\begin{Schunk}
\begin{Sinput}
R> library(confintROB)
R> data(medication)
R> str(medication)
\end{Sinput}
\begin{Soutput}
'data.frame':	1242 obs. of  5 variables:
 $ obs  : int  1 2 3 4 5 6 7 8 9 10 ...
 $ id   : int  1 1 1 1 1 1 1 1 1 1 ...
 $ treat: int  1 1 1 1 1 1 1 1 1 1 ...
 $ time : num  0 0.333 0.667 1 1.333 ...
 $ pos  : num  107 100 100 100 100 ...
\end{Soutput}
\end{Schunk}
where \code{obs} denotes the observation number, \code{id} identifies the 64 subjects of the sample , \code{treat} is a dichotomous variable to define control (\code{treat=0} for 27 subjects) and treatment (\code{treat=1} for 37 subjects), \code{time} takes 21 values (from 0 to 6.67 increasing by 0.33 at each time point), \code{pos} is the positive mood score, varying from 100 to 500 (mean = 167.79). Although the data set does not contain any \texttt{NA}s, it is unbalanced because 33 subjects were assessed fewer than 21 times, which amounts to 7.59\% of data missing.

\begin{figure}[ht]
\centering
\includegraphics{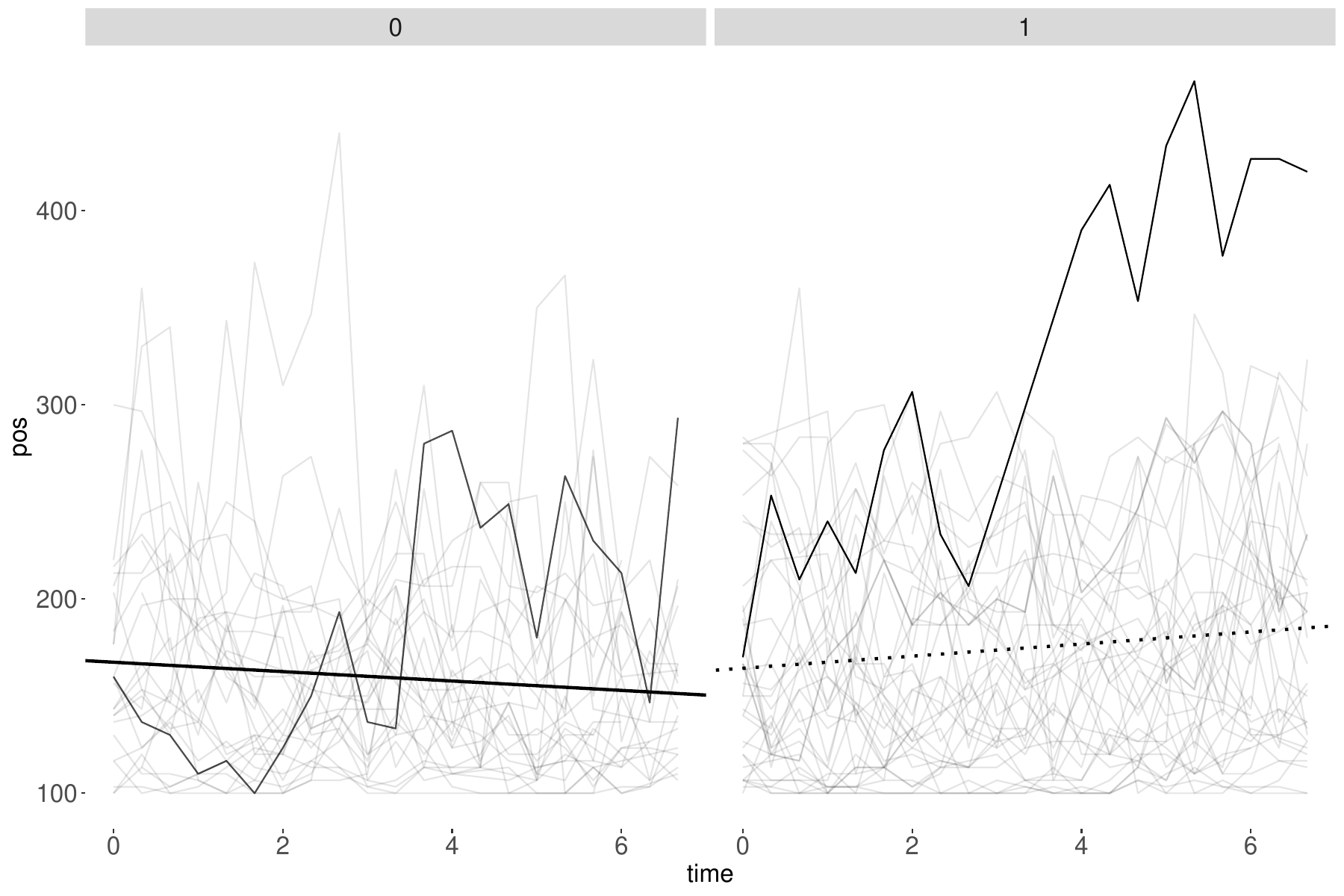}
\caption{\footnotesize The individual positive mood trajectories (\code{pos}) as a function of waves of assessment (\code{time}) and treatment (\code{treat = 0} on the left and \code{treat = 1} on the right panel). The regression lines are based on the Maximum Likelihood (ML) estimates of the model in Equation~\eqref{LMM_eq}. The solid black line is the predicted average trajectory for individuals of \code{treat = 0} and the dotted line is the predicted average trajectory for individuals of \code{treat = 1}. The lineplots represent the individual trajectories (The clearer the lines, the closer the individual estimated weights (by RSE) to 1.).}
\label{fig:medicationDATA}
\end{figure}

The group-specific representation is presented in Figure~\ref{fig:medicationDATA}. Here, we see each subject's measurements represented by lineplots. We can see that individual trajectories are extremely variable between subjects (between-subject variability) and unstable in time (within-subject variability). If we try to represent the trajectories by straight lines, the two groups seem to differ with respect to their slopes, but not on their starting levels. To emphasize this, we plotted with a solid (\code{treat = 0}) and a dotted (\code{treat = 1}) line the group-implied average straight-line trajectories. The shade of the individuals' lineplots provides information about the weight(s) assigned to each individual by the estimator. The lower the weight(s) of an individual, the darker their line, thus the greater the chance that the individual be considered a between-subject outlier \citep{Koller2016}. Two individual trajectories are particularly darker, because those subjects' data appear extreme with respect to their group, probably due to their extremely high slope scores \citep{Koller2016}. Their weights are 0.41 and 0.12, for the individual in the \code{treat = 0} and \code{treat = 1} group, respectively. A second individual in the \code{treat = 1} group has a weight of 0.90, rendering their line less distinguishable than those of the two individuals mentioned before. According to \cite{mason2024}, these are the outliers that have the greatest impact on the estimates of the time by group interaction fixed parameter, which is the parameter of main interest in many group comparison studies, because it captures the different effect of time across the two groups (e.g., differential treatment effect). The inferential results of this particular parameter are also greatly affected by this type of outlier. The presence of random slope between-subject outliers can greatly affect the rejection rates regarding the fixed interaction effect \cite{mason2024}.\\ 

A first step towards assessing whether outliers influenced estimation consists in comparing parameter estimates from a classical estimator, here ML, with those from a robust alternative, here RSE. In this example, we are particularly motivated by the effect of \code{time} on \code{pos} possibly moderated by \code{treat}. Below, we show the code to model this hypothesis taking into account that participants may differ in their intercept and slope values.
\begin{Schunk}
\begin{Sinput}
R> library(lmerTest)
R> model.ML <- lmer(pos ~ treat * time + (time|id), data = medication, 
+                   REML = "FALSE")
\end{Sinput}
\end{Schunk}
The object \code{model.ML} contains the parameter estimates obtained with the \fct{lmer} function based on the ML estimator because of the argument \code{REML=FALSE} (for more details about this function see \cite{bates2015lme4}). To obtain the equivalent model with RSE estimates, we use the function \fct{rlmer}. 
\begin{Schunk}
\begin{Sinput}
R> library(robustlmm)
R> model.RSE <- rlmer(pos ~ treat * time + (time|id), data = medication, 
+                     method = "DAStau", init = inits.RSE)
\end{Sinput}
\end{Schunk}
An easy way to compare estimates is to produce a table using the \fct{compare} function from package \pkg{robustlmm} calling objects \code{model.ML} and \code{model.RSE}:
\begin{Schunk}
\begin{Sinput}
R> compare(model.ML, model.RSE)
\end{Sinput}
\end{Schunk}
\begin{table}[ht]
\centering
\begin{tabular}{rll}
  \hline
  & model.ML & model.RSE \\ 
  \hline
Coefficients (Std. Error) &  &  \\ 
  (Intercept) & 167.46 ( 9.33)  & 163.831 ( 9.77) \\ 
  treat &  -3.11 (12.33)  &   0.232 (12.90) \\ 
  time &  -2.42 ( 1.73)  &  -2.551 ( 1.40) \\ 
  treat:time &   5.54 ( 2.28)  &   4.206 ( 1.85) \\ 
   \hline
Variance components &  &  \\ 
  (Intercept) $|$ id & 45.95 & 48.51 \\ 
  time $|$ id &  7.98 &  6.38 \\ 
   \hline
Correlations &  &  \\ 
  (Intercept) $\times$ time $|$ id & -0.332 & -0.431 \\ 
   \hline
$\sigma$ & 35.1 & 27.8 \\ 
   \hline
deviance & 12680 &  \\ 
  rho.e &  & smoothed Huber (k = 1.345, s = 10) \\ 
  rho.$\sigma$.e &  & smoothed Huber, Proposal 2 (k = 1.345, s = 10) \\ 
  rho.b\_1 &  & smoothed Huber (k = 5.14, s = 10) \\ 
  rho.$\sigma$.b\_1 &  & smoothed Huber (k = 5.14, s = 10) \\ 
   \hline
\end{tabular}
\caption{Comparison of classical and robust fits. Rendering of output of \code{xtable(compare(model.ML,model.RSE))}.} 
\label{tab:compareMethods}
\end{table}%
Table~\ref{tab:compareMethods} shows the output of \code{xtable} in rendered form. The \emph{Coefficients} section contains the estimates of the fixed parameters, the estimated variance of the random effects are in the section \emph{Variance components}, their estimated correlation in the \emph{Correlations} section, and the standard deviation of error terms is called $\sigma$. The last part of the output gives the deviance for \code{model.ML} and the functions used in the algorithm to obtain the robust estimates (see \cite{Koller2016} for more details). The estimates of the variance components and the fixed-effect coefficients for the intercept, \code{treat} and \code{time} are at times very similar (e.g., fixed and random effects of Intercept) and at times different (e.g., fixed and random effects of time) across the two fits. A meaningful difference in estimates is the fixed effect of the interaction, which is slightly smaller with RSE. This is coherent with the fact that the very high slope values of two outliers in group \code{treat=1} tend to increase the difference in the average slope between the two groups. These results can be complemented by CIs, which are also useful for inferential decisions. We therefore introduce the \fct{confintROB} function that applies the percentile wild bootstrap to the model estimated by ML. 
\begin{Schunk}
\begin{Sinput}
R> set.seed(3)
R> wild.ML <- confintROB(object = model.ML, boot.type = "wild",
+                        verify.saved = wild.ML)
\end{Sinput}
\end{Schunk}
The function \fct{confintROB} can be applied to any object of class \class{lmerMod}, \class{lmerModLmerTest}, \class{rlmerMod} or \class{varComprob}. In this example, \code{model.ML} is an object of class \class{lmerMod} (see Table~\ref{table_ci}). The percentile bootstrap CI being the default argument, we specify that we want the wild bootstrap with the \code{boot.type} argument. We use the default values for argument \code{nsim} and the argument \code{level} (respectively, 5000 bootstrap samples and .95 confidence level).
\begin{Schunk}
\begin{Sinput}
R> wild.ML
\end{Sinput}
\begin{Soutput}
                                2.5 %       97.5 %
(Intercept)               150.1803810 185.45316421
treat                     -26.9182945  20.44265435
time                       -5.3947424   0.67222061
treat:time                  1.3825904   9.89946119
Sigma id (Intercept)       37.1847684  53.02209619
Sigma id time               5.2272994  10.76968251
Sigma id (Intercept) time  -0.5461602  -0.07403813
Sigma Residual             28.7413919  41.48149925
attr(,"fullResults")
  Full results of confintROB, a list with components:
   "Percentile", "bootstrap_estimates" 
\end{Soutput}
\end{Schunk}
Results are saved in the object \code{wild.ML}. With the \fct{print} function, the matrix of selected CIs is returned. In this example, the CIs for all the parameters are provided (the default). If the interest is on a subset of the parameters, the argument \code{parm} can be used. For instance, to obtain the CI for the variable \code{treat} only, one would specify \code{parm = 2} or \code{parm = "treat"}. To obtain the CIs for \code{treat} and \code{Sigma id time}, one would specify \code{parm = c(2, 6)} or \code{parm = c("treat", "Sigma id time")}. The first column of the output matrix always contains the names of the parameters. When the name begins with \code{Sigma}, the corresponding CIs are for the standard deviations of residuals (\code{Sigma Residual}), random effect, or correlation between random effects. For random effects, the string \code{Sigma} is followed by the name of the cluster (here, \code{id} \code{(Intercept)} indicates that it is the random effect of the intercept, when random slopes are supposed, the name of the variable follows \code{id} (here, \code{time}) and correlation is indicated with both names (in this example, \code{(Intercept) time}). The remaining effects are fixed. The second and third columns represent, respectively, the lower and upper bound of the CIs. The string \code{attr(,"fullResults")} indicates that the object \code{wild.ML} also contains an attribute of class \class{fullResults}, which is a list containing the complete matrix with the percentile version of the CIs (\code{"Percentile"}) and all the parameter estimates of the simulated samples (\code{"bootstrap\_estimates"}).
We now apply the wild bootstrap to RSE estimates by replacing the \code{model.ML} object by \code{model.RSE}.
\begin{Schunk}
\begin{Sinput}
R> set.seed(3)
R> wild.RSE <- confintROB(object = model.RSE,
+                         boot.type = "wild",
+                         verify.saved = wild.RSE)
R> wild.RSE
\end{Sinput}
\begin{Soutput}
                                 2.5 %       97.5 %
(Intercept)               146.44579838 181.89124806
treat                     -23.80293651  23.55395617
time                       -5.50695767   0.57216356
treat:time                  0.02290454   8.63920558
Sigma id (Intercept)       37.14917091  53.38396261
Sigma id time               5.22385401  10.89180572
Sigma id (Intercept) time  -0.54931053  -0.07081393
Sigma Residual             28.69255451  41.42750808
attr(,"fullResults")
  Full results of confintROB, a list with components:
   "Percentile", "bootstrap_estimates" 
\end{Soutput}
\end{Schunk}
Again, estimates are extremely similar to the ones obtained with ML, with a slight negative shift in the CIs for the fixed-effect parameters of the \code{Intercept}, \code{treat}, and the interaction between \code{treat} and \code{time}. 

Both classical and robust estimators agree in concluding that there is no effect of \code{treat}, \code{time}, and a moderately strong, negative correlation between random effects in the original sample. However, whereas the classical estimator clearly concludes that there is a time by group interaction effect, the conclusion drawn from the robust estimator is not as clearcut, with a lower bound much closer to 0. In substantive terms, this difference in results is of utmost importance, as it may potentially invalidate a treatment.

\begin{figure}[h]
\centering
\includegraphics{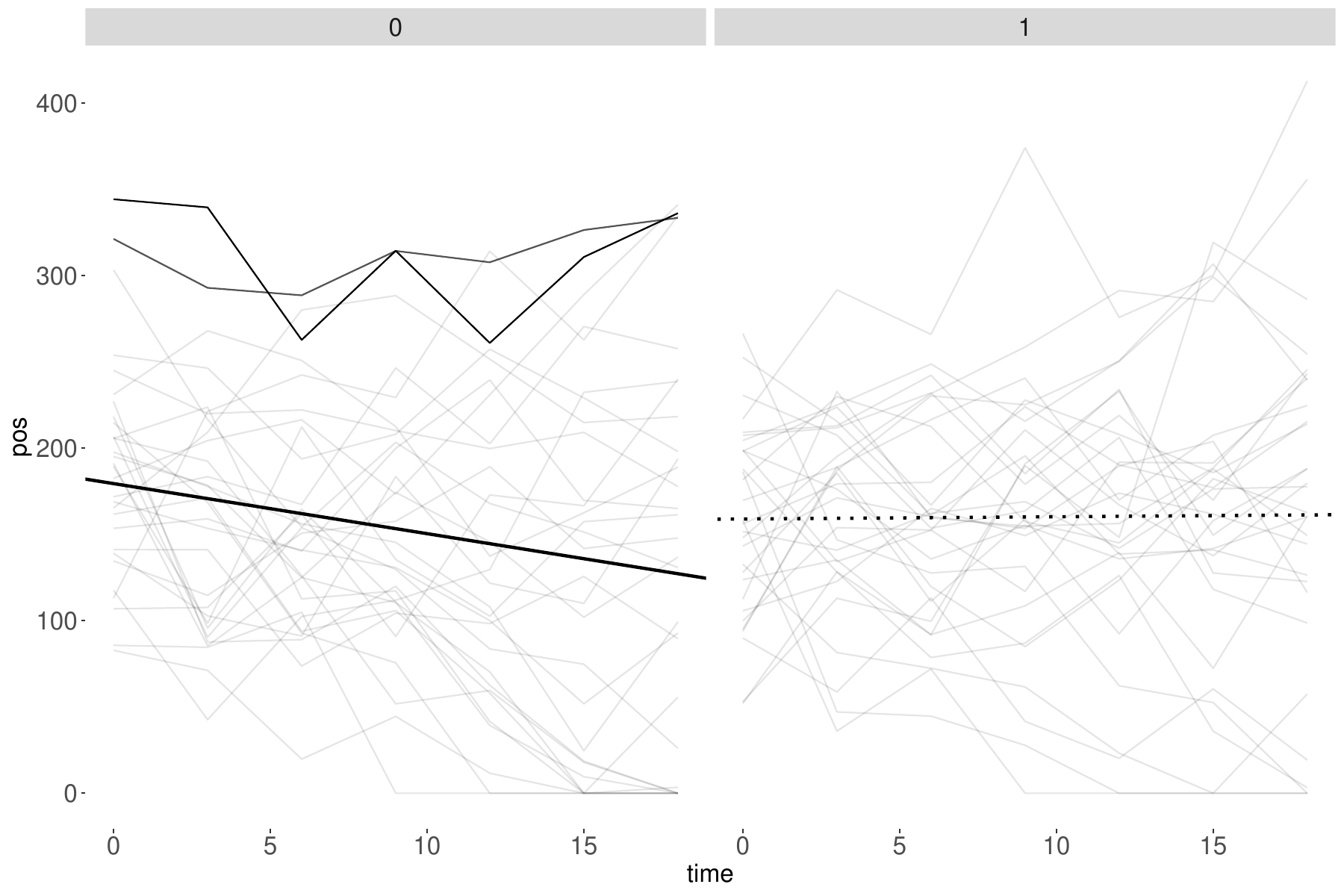}
\caption{\footnotesize The individual positive mood trajectories (\code{pos}) as a function of waves of assessment (\code{time}) and treatment (\code{treat = 0} on the left and \code{treat = 1} on the right panel). The regression lines are based on the Maximum Likelihood (ML) estimates of the model in Equation~\eqref{LMM_eq}. The solid black line is the predicted average trajectory for individuals of \code{treat = 0} and the dotted line is the predicted average trajectory for individuals of \code{treat = 1}. The lineplots represent the individual trajectories (The clearer the lines, the closer the individual estimated weights (by RSE) are to 1.).}
\label{medsimDATA}
\end{figure}

Obviously, in this first example with real data it is impossible to know the true values of the parameters in the population, and some may suspect that the robust estimator is simply more conservative or less powerful than the classical estimator. Therefore, we introduce a second example (\code{medsim}) inspired by the first one, but based on simulated data, for which we know the true parameter values in model \eqref{LMM_eq} and \eqref{Nassump}: $\boldsymbol{\gamma}=(167.46, ~-3.11, ~-2.42, ~4.00)$, $\boldsymbol{\theta} = (2111.54, ~63.74, ~-121.63)^T$ and $\sigma_{\varepsilon}^2=1229.93$. 
\begin{Schunk}
\begin{Sinput}
R> str(medsim)
\end{Sinput}
\begin{Soutput}
'data.frame':	420 obs. of  5 variables:
 $ obs  : int  1 2 3 4 5 6 7 8 9 10 ...
 $ id   : int  1 1 1 1 1 1 1 2 2 2 ...
 $ time : int  0 3 6 9 12 15 18 0 3 6 ...
 $ treat: int  0 0 0 0 0 0 0 0 0 0 ...
 $ pos  : num  205.8 192.5 126 51.8 59.6 ...
\end{Soutput}
\end{Schunk}
The sample size is similar to that of the real data example ($N=60$) but with balanced and complete data and with fewer repeated assessments per subjects ($J_i=J=7$), to reduce the file size. Time is equal to (\code{0,3,6,9,12,15 and 18}). Two participants from to the control group were originally in the treatment group: we intentionally changed their respective label to create outliers (see the dark trajectories in the left panel of the Figure~\ref{medsimDATA}). Unlike the ``outliers'' in (\code{treat = 1}) highlighted in Figure~\ref{fig:medicationDATA}, which tended to accentuate the difference in trajectories between participants in the two groups and thereby increase the probability of obtaining a significant effect for the interaction parameter $\gamma_3$, these outliers should increase the confusion between the two groups and, in this case, decrease the probability of rejecting the related null hypothesis (similarly to the outlier in (\code{treat = 0}) highlighted in Figure~\ref{fig:medicationDATA}).
After estimating the same model on the simulated data by creating objects  \code{modelSim.ML} and \code{modelSim.RSE} respectively, the estimation of the parameters can once again be directly compared using the \fct{compare} function:
\begin{Schunk}
\begin{Sinput}
R> modelSim.ML <- lmer(pos ~ treat * time + (time|id), data = medsim, REML = "FALSE")
\end{Sinput}
\end{Schunk}
\begin{Schunk}
\begin{Sinput}
R> modelSim.RSE <- rlmer(pos ~ treat * time + (time|id), data = medsim, 
+                        method = "DAStau", init = inits.SimRSE)
\end{Sinput}
\end{Schunk}
\begin{Schunk}
\begin{Sinput}
R> compare(modelSim.ML, modelSim.RSE)
\end{Sinput}
\end{Schunk}
\begin{table}[ht]
\centering
\begin{tabular}{rll}
  \hline
  & modelSim.ML & modelSim.RSE \\ 
  \hline
Coefficients (Std. Error) &  &  \\ 
  (Intercept) & 179.50 (10.18) & 178.55 (10.64) \\ 
  treat & -20.54 (14.40) & -21.07 (15.04) \\ 
  time &  -2.90 ( 1.14) &  -3.00 ( 1.21) \\ 
  treat:time &   3.03 ( 1.61) &   3.19 ( 1.71) \\ 
   \hline
Variance components &  &  \\ 
  (Intercept) $|$ id & 51.13 & 52.83 \\ 
  time $|$ id &  5.90 &  6.17 \\ 
   \hline
Correlations &  &  \\ 
  (Intercept) $\times$ time $|$ id & -0.234 & -0.248 \\ 
   \hline
$\sigma$ & 32.6 & 32.8 \\ 
   \hline
deviance & 4428 &  \\ 
  rho.e &  & smoothed Huber (k = 1.345, s = 10) \\ 
  rho.$\sigma$.e &  & smoothed Huber, Proposal 2 (k = 1.345, s = 10) \\ 
  rho.b\_1 &  & smoothed Huber (k = 5.14, s = 10) \\ 
  rho.$\sigma$.b\_1 &  & smoothed Huber (k = 5.14, s = 10) \\ 
   \hline
\end{tabular}
\caption{Comparison of classical and robust fits on the simulated dataset. Rendering of output, \code{xtable(compare(modelSim.ML, modelSim.RSE))}.} 
\label{tab:compareSimdata}
\end{table}%
Again, the parameter estimates for fixed effects, their standard errors, and variance components are very similar for both estimators. By applying the wild bootstrap using the \fct{confintROB} function to these two new objects, we once again obtain a CI for each parameter of the model.
\begin{Schunk}
\begin{Sinput}
R> set.seed(3)
R> wildSim.ML <- confintROB(object = modelSim.ML,
+                           boot.type = "wild",
+                           verify.saved = wildSim.ML)
R> wildSim.ML
\end{Sinput}
\begin{Soutput}
                                2.5 %       97.5 %
(Intercept)               158.5044035 201.29909601
treat                     -48.8550232   6.93072963
time                       -5.2582788  -0.51693479
treat:time                 -0.1187373   6.17566692
Sigma id (Intercept)       38.4990062  62.30562438
Sigma id time               4.6069444   7.04286448
Sigma id (Intercept) time  -0.4960982   0.05952634
Sigma Residual             28.0164266  37.43709604
attr(,"fullResults")
  Full results of confintROB, a list with components:
   "Percentile", "bootstrap_estimates" 
\end{Soutput}
\end{Schunk}
\begin{Schunk}
\begin{Sinput}
R> set.seed(3)
R> wildSim.RSE <- confintROB(object = modelSim.RSE,
+                            boot.type = "wild",
+                            verify.saved = wildSim.RSE)
R> wildSim.RSE
\end{Sinput}
\begin{Soutput}
                                 2.5 %       97.5 %
(Intercept)               157.54765680 199.99603194
treat                     -49.50558142   6.54352734
time                       -5.33278702  -0.59838226
treat:time                  0.04245368   6.33599786
Sigma id (Intercept)       38.49685977  62.11219643
Sigma id time               4.60446985   7.04626841
Sigma id (Intercept) time  -0.49917653   0.06116436
Sigma Residual             27.94773579  37.41651808
attr(,"fullResults")
  Full results of confintROB, a list with components:
   "Percentile", "bootstrap_estimates" 
\end{Soutput}
\end{Schunk}
Once again, the confidence intervals (CIs) are very similar for each parameter, and again, the interaction effect is marginally significant only with the robust estimator, thus leading to an opposite substantive finding concerning the treatment. This result suggests that it is not a matter of a more conservative or less powerful method, but rather the nature of the outliers that determines the outcome of the inferential analysis.

\section{Summary and discussion} \label{sec:summary}
The main purpose of the current package is to provide a complementary tool to the robust estimators of LMM obtained with the packages \pkg{robustlmm} and \pkg{robustvarComp}, so as to facilitate comparisons with classical estimators, and to allow for inferential decisions based on CIs for both fixed and variance components. The output and options are constructed to be as similar as possible to the function \fct{confint} from package \pkg{stats} and \fct{confint.merMod} from package \pkg{lme4} so that users of those packages can readily become accustomed to results from the \pkg{confintROB} package.

In the future, it would be worthwhile to investigate also non-parametric tests. While parametric tests are more powerful when the assumptions are met, these assumptions can be quite demanding, and in cases where they are not met, parametric tests are generally less effective \citep{anderson1961}. In the context of the LMM, the cluster bootstrap is available for classical estimator in \proglang{R} through the \pkg{ClusterBootstrap} package \citep{deen2019}. Unfortunately, its performance in terms of coverage rates was found to be inferior to the parametric and the wild bootstrap CIs \citep{Modugno2013}. Moreover, with contaminated data, the proportion of outliers can be very high in the bootstrap samples, thus biasing parameter estimation \citep{salibian2006, salibian2008}.

Permutation tests are also non-parametric alternatives that, like the wild bootstrap, have the advantage, compared to the cluster bootstrap, of maintaining the same proportion of outliers in the resamples as in the original sample. In permutation tests, the same observations are retained but with permuted characteristic, such as the group labels in a two-sample independent t-test or the predictor-response pairs in simple linear regressions, as if the link between the predictor and the response variable was null, to obtain an empirical distribution of the parameter (or statistic) of interest under the null hypothesis. Despite the complexity of the repeated measures data structure, permutation methods have been developed for classical LMMs and made accessible, for example, through the \code{lmm.perm} function in the \pkg{minque} package in \proglang{R} \citep{wu2019}. These tests are well-suited for group comparisons \cite{shields2018}. Under the null hypothesis, individuals from both groups come from the same population and are therefore independent and identically distributed. It suffices to permute the group labels. Several permutation methods for LMMs are based on the likelihood of the model and are thus not applicable to robust estimators \citep[e.g.,][]{krzciuk2014, lee2012, rao2019}. In their first study, \cite{krzciuk2014} also combined the Wald statistic with the permutation method. The Wald statistic was computed in the original sample and in the permuted samples. The $p$-value was defined as the proportion of Wald statistic values obtained from the permuted samples that were greater than the one obtained from the original sample. However, the results regarding the variances were significantly poorer than the two other methods based on likelihood models, so this method was discarded. Nevertheless, Krzciuk's method appears to be a good alternative to tests of fixed effects with robust estimators, especially considering the excellent performance of the original test with this class of estimators.

We hope that with the \pkg{confintROB} package, LMM users who are wary of outliers can have the best of both worlds: a robust estimation procedure for their LMM estimates (via the \pkg{robustlmm} or \pkg{robustvarComp} packages) coupled with state-of-the-art confidence intervals, as currently implemented in the \pkg{stats} and the \pkg{lme4} packages for classical estimators. We believe that being able to compare parameter estimates as well as their CIs across classical and robust estimators is a valuable tool when testing research hypotheses that could easily go astray because of a few, odd, and yet influential observations.


\section*{Computational details}

The results in this paper were obtained using
\proglang{R}~4.3.1 with the
\pkg{robustlmm}~3.2-3 and \pkg{lme4}~1.1-34 packages. \proglang{R} itself
and all packages used are available from the Comprehensive
\proglang{R} Archive Network (CRAN) at
\url{https://CRAN.R-project.org/}.

\section*{Acknowledgments}

We thank Drs. Lucia Modugno and Simone Giannerini for helping us with their \proglang{R} script to compute wild bootstraps.


\bibliography{refs}

\newpage

\begin{appendix}

\end{appendix}


\end{document}